\documentclass[atoms,article,submit,moreauthors]{Definitions/mdpi}
\firstpage{1}
\pubvolume{0}
\issuenum{0}
\articlenumber{0}
\pubyear{2026}
\copyrightyear{2026}
\datereceived{ }
\daterevised{ }
\dateaccepted{ }
\datepublished{ }
\usepackage{bm}
\usepackage{mathtools}
\usepackage{siunitx}
\usepackage{needspace}
\newcommand{\SO}{\mathrm{SO}}
\newcommand{\NR}{\mathrm{NR}}
\newcommand{\code}[1]{\texttt{\detokenize{#1}}}

\Title{Multiconfigurational and Spin--Orbit Extensions of the \mbox{BSR--RMT} Interface}

\Author{Aaron T. Bondy $^{1,*}$, Juan C. Del Valle $^{2}$,
Kathryn R. Hamilton $^{3}$,
and Klaus Bartschat $^{1}$}

\AuthorNames{Aaron T. Bondy, Juan C. del Valle,
Kathryn R. Hamilton, and Klaus Bartschat}

\address{%
$^{1}$ \quad Department of Physics and Astronomy, Drake University,
Des Moines, IA 50311, USA\\
$^{2}$ \quad Instituto de F\'isica, Universidad Nacional
Aut\'onoma de M\'exico, Ciudad de M\'exico 04510, Mexico\\
$^{3}$ \quad Department of Physics, University of Colorado Denver,
Denver, CO 80204, USA}
\corres{Correspondence: aaron.bondy@drake.edu}

\abstract{We report the extension of the B-spline R-matrix (BSR) to R-matrix with time dependence (RMT) interface to include support for multiconfiguration target states and \texorpdfstring{$jK$}{jK} coupling, including the one-body spin--orbit (SO) interaction.
Inner-region input data generated by BSR are transformed through the interface for time propagation in RMT.
Two examples demonstrate the working interface: high-order harmonic generation in helium and weak-field, near-resonance response in carbon.
The helium comparisons employ previously studied laser parameters, displaying excellent agreement with the established Belfast \hbox{R-matrix} input codes in the time-dependent dipole and harmonic spectrum, both with and without SO.
Further structure comparisons in helium show close agreement in the SO-induced changes in energy and oscillator strengths.
The two codes coincide on the much larger SO-induced change in the dipole response in carbon, demonstrating agreement for multiconfigurational target states.
These extensions allow the flexibility of BSR structure calculations to be used in time-dependent studies of heavier and more complex atoms.
}

\let\linenumbers\nolinenumbers
\begin{document}

\section{Introduction}\label{sec:introduction}
To characterize the response of multielectron atoms to laser fields, both atomic structure and dynamics must be well understood~\cite{KrauszIvanov2009}.
Exchange, electron correlation, and relativistic effects determine energy levels and transition amplitudes, further shaping excitation, ionization, and the ensuing dynamics.
A multitarget and multiconfiguration expansion of the ionic target facilitates the inclusion of correlation in the atomic response.
Reconstruction and control of a time-dependent two-electron wave packet has been demonstrated in helium~\cite{Ott2014}, and the coherent motion of valence electrons in krypton ions has been observed~\cite{Goulielmakis2010}. 
Further, differences in photoionization delays associated with the spin--orbit-split $4\mathrm{d}$ channels in xenon have been observed~\cite{Zhong2020}.
The spin--orbit (SO) interaction splits atomic terms into fine-structure levels and can mix terms with different multiplicity but common total electronic angular momentum and parity, thereby creating new excitation and ionization pathways and dynamics absent in the nonrelativistic case.

The first version of time-dependent R-matrix theory was introduced by Burke and Burke~\cite{BurkeBurke1997}.
The combination in time-dependent calculations of an inner-region eigenstate basis with an outer-region finite-difference grid was first developed for one-electron systems~\cite{Nikolopoulos2008}.
The R-matrix with time dependence (RMT) method extended this treatment to multielectron atoms~\cite{Moore2011}.
A later extension of RMT enabled the treatment of interaction with arbitrarily polarized laser fields~\cite{Clarke2018,Brown2020}.
In the standard RMT pathway, the input (inner region structure data) is produced by the Belfast R-matrix codes: RMII~\cite{Burke1994} for nonrelativistic calculations and RMI~\cite{Berrington1995}, modified for use with RMT~\cite{Wragg2020}, for calculations using $jK$ coupling and those including SO.
RMT has demonstrated how the SO interaction affects the excitation pathways in krypton driven by time-delayed parallel and perpendicular pulses~\cite{Wragg2019}.

The B-spline R-matrix (BSR) method~\cite{Zatsarinny2006} provides an alternative method for preparing inner-region input, allowing a flexible treatment of complex systems via multiconfigurational states constructed from term-dependent, nonorthogonal orbital sets~\cite{Zatsarinny1996}.
Schneider, Hamilton, and Bartschat~\cite{Schneider2022} advocated an interface between BSR and RMT to take advantage of these features of BSR.
The \mbox{BSR--RMT} interface developed by del Valle et al.~\cite{DelValle2025} realized this possibility within $LS$ coupling using single-configuration target states.
A key issue, beyond the interface routines themselves, was converting the Condon--Shortley (CS) convention employed in BSR for the electron--electron interaction and the associated angular changes to the Fano--Racah (FR) treatment~\cite{FanoRacah1959} used in the Belfast codes, including RMT.

The present work extends the \mbox{BSR--RMT} interface to multiconfiguration targets and $jK$ coupling, including treatment of the one-body SO interaction.
A recent related study~\cite{BondyBartschat2026} interfaced semirelativistic BSR inner region data with the asymptotic programs STGF and STGBF~\cite{Badnell2018} for the stationary photoionization of highly charged two-electron ions.
As in the first \mbox{BSR--RMT} implementation, we have constructed interface utilities that generate the input files required by RMT, without making any changes to RMT itself.

The following Section~\ref{sec:method} provides a brief theoretical overview of the multiconfiguration target states, channel coupling, and Hamiltonian used in the calculations, followed by a description of the interface utilities in Section~\ref{sec:interface}.
In Section~\ref{sec:results}, we test the interface by conducting comparisons with the established \mbox{RMII--RMT} and \mbox{RMI--RMT} pathways~\cite{Wragg2020} for high-order harmonic generation (HHG) of helium and the weak-field, near-resonance, long post-pulse dynamics in carbon.
For these test cases, which are tractable with relatively small computational resources, the dynamic RMT results obtained with the BSR input are in excellent agreement with those using input from the Belfast codes.
Section~\ref{sec:conclusions} summarizes and interprets the results with a discussion of the many applications to more complex systems made possible by the extension to \mbox{BSR--RMT} described herein.

\section{Theory and Computational Method}\label{sec:method}
\subsection{Multiconfiguration target states}\label{sec:target-states}
The residual target ion has $N$ electrons, while the inner region structure data are computed for the ($N+1$)-electron system whose wave function is propagated in RMT.
A target state with total angular momentum $J_t$ and parity $\pi_t$ is expanded in configuration-state functions (CSFs) as
\begin{equation}
 \Phi_{\gamma}^{J_t\pi_t}
 =\sum_{\nu}c_{\gamma\nu}\varphi_{\gamma\nu}^{J_t\pi_t}.
 \label{eq:target}
\end{equation}
Here, $\gamma$ labels distinct target states with the same $J_t$ and $\pi_t$, and $\nu$ denotes the CSFs used to expand state $\gamma$.
The CSFs can be chosen separately for each state; however, when a common set is used, the coefficients $c_{\gamma\nu}$ depend on the state.
Within a CSF, the total angular momentum is the vector addition of the total orbital and total spin angular momenta, $\bm{J}_t=\bm{L}_t+\bm{S}_t$.
With SO included, target states can contain CSFs with different $L_t$ and $S_t$, which are coupled to the same $J_t$.
In BSR, target states are provided directly from the multiconfiguration Hartree--Fock (MCHF) code~\cite{FroeseFischer2007,ZatsarinnyFischer2009}.
Here, we use a slightly modified version of MCHF where only the spin--orbit contribution can be selected in order to compare with \mbox{RMI--RMT}, as described in Section~\ref{sec:hamiltonian}.

MCHF also uses the CS phase convention. Thus, the coefficients of the CSFs for a multiconfigurational target state from MCHF must be converted to the FR convention required here.
As described in Ref.~\cite{DelValle2025}, the spherical harmonics in the two conventions are related by
\begin{equation}
 Y_{\ell m}^{\mathrm{FR}}=\mathrm{i}^{\ell}Y_{\ell m}^{\mathrm{CS}}.
 \label{eq:orbital-phase}
\end{equation}
In a particular target state, let $n_{\nu q}$ be the number of electrons in subshell $q$ of CSF $\nu$, and let $\ell_q$ be the orbital angular momentum of that subshell.
The product of the single-electron phases is $\mathrm{i}^{P_\nu}$, where
\begin{equation}
 P_\nu=\sum_q n_{\nu q}\ell_q.
 \label{eq:occupation-phase}
\end{equation}
Suppressing the $J_t\pi_t$ label common to all CSFs, the CS-to-FR transformation required is thus
\begin{equation}
 \varphi_{\gamma\nu}^{\mathrm{FR}}
 =\mathrm{i}^{P_\nu}\varphi_{\gamma\nu}^{\mathrm{CS}}.
 \label{eq:csf-phase}
\end{equation}
The coefficient of each CSF in Eq.~\eqref{eq:target} needs to be multiplied by $\mathrm{i}^{-P_\nu}$ to leave the target state unmodified.
The overall phase of the entire state is arbitrary.
Thus, if one CSF is taken as a reference with a phase factor $\mathrm{i}^{P_{\mathrm{ref}}}$, all transformed coefficients are multiplied by $\mathrm{i}^{P_{\mathrm{ref}}}$, giving
\begin{equation}
 c_{\gamma\nu}^{\mathrm{FR}}
 =\mathrm{i}^{P_{\mathrm{ref}}-P_\nu}c_{\gamma\nu}^{\mathrm{CS}}
 =(-1)^{(P_{\mathrm{ref}}-P_\nu)/2}c_{\gamma\nu}^{\mathrm{CS}}.
 \label{eq:coefficient-phase}
\end{equation}

\noindent Since all CSFs within a state have the same parity, $P_{\mathrm{ref}}-P_\nu$ is even and the conversion requires at most a change of sign.
For example, the carbon target configurations $1\mathrm{s}^22\mathrm{s}^22\mathrm{p}$ and $1\mathrm{s}^22\mathrm{p}^3$ have $P_\nu=1$ and $3$, respectively.
If we choose the first as the reference, then the sign of the second must flip.
For single-configuration target states, the phase is arbitrary and no changes are required.

\subsection{Channel coupling and inner-region expansion}\label{sec:channels}
The target states in Eq.~\eqref{eq:target} are coupled to an additional electron to form the $(N+1)$-electron channel functions.
In $LS$ coupling, $\bm{L}=\bm{L}_t+\bm{\ell}$ and $\bm{S}=\bm{S}_t+\bm{s}$, with $s=1/2$, are formed separately, and a channel is characterized by its $LS\pi$ symmetry.
Alternatively, in $jK$ coupling, $\bm{J}_t$ is first coupled to the additional electron's orbital angular momentum $\bm{\ell}$ to form $\bm{K}$.
After that, its spin $s=1/2$ is coupled to $\bm{K}$ to produce the total electronic angular momentum $J$~\cite{Edmonds1996}:
\begin{equation}
 \left[\left[(L_tS_t)J_t,\ell\right]K,s\right]J,
 \qquad \pi=\pi_t(-1)^\ell.
 \label{eq:jk}
\end{equation}
The parity is obtained by multiplying the target and additional electron parities.
In a target containing SO-mixing, this coupling applies separately to each $LS$ component of its expansion in Eq.~\eqref{eq:target}.

In the R-matrix expansion~\cite{Burke2011}, the BSR method~\cite{Zatsarinny2006} represents an inner-region eigenstate as
\begin{equation}
 \Psi_k^{J\pi}
 =\mathcal{A}\sum_{ij}\overline{\Phi}_i^{J\pi}
   \frac{B_j(r)}{r}a_{ijk}
  +\sum_q\chi_q^{J\pi}b_{qk}.
 \label{eq:inner}
\end{equation}
Here, $\mathcal{A}$ is the antisymmetrization operator.
The channel function $\overline{\Phi}_i^{J\pi}$ combines a target state from Eq.~\eqref{eq:target} with the angular and spin functions of the additional electron, coupled according to Eq.~\eqref{eq:jk}.
The radial coordinate is $r$, and B-splines $B_j(r)$ are employed to form the radial basis within the R-matrix boundary~\cite{Zatsarinny2006,Bachau2001}.
The indices $i$, $j$, and $k$ label channels, \hbox{B-splines}, and inner-region eigenstates, respectively.
$\chi_q^{J\pi}$  are square-integrable functions describing additional short-range correlation, including bound-orbital configurations absent from the channel expansion.
The coefficients $a_{ijk}$ and $b_{qk}$ result from diagonalizing the inner-region Hamiltonian in this basis.

The Belfast R-matrix reference calculations use RMII~\cite{Burke1994} in $LS$ coupling and RMI~\hbox{\cite{Berrington1995,Wragg2020}} in $jK$ to generate the inner-region atomic structure used as input for RMT.
These results are then compared to BSR input data while keeping a consistent atomic description in both cases.

\subsection{Hamiltonian used in the comparisons}\label{sec:hamiltonian}
In the SO comparisons, the nonrelativistic Hamiltonian is supplemented with the one-body spin--orbit operator.
For the inner-region ($N+1$)-electron atomic calculation (in atomic units)
\begin{equation}
 H_{\mathrm{atom}}=H_{\mathrm{NR}}+H_{\mathrm{SO}},
 \qquad
 H_{\mathrm{SO}}=
 \frac{\alpha^2 Z}{2}\sum_{i=1}^{N+1}
 \frac{\bm{\ell}_i\!\cdot\!\bm{s}_i}{r_i^3},
 \label{eq:hamiltonian}
\end{equation}
where $Z$ is the nuclear charge and $\alpha \approx 1/137$ is the fine-structure constant.
The target description uses the same operator in the $N$-electron case.
Of course, the NR calculations omit the SO term. 

In BSR, the effective SO interaction by default includes a contribution from closed-shell electrons beyond simply the nuclear term in Eq.~\eqref{eq:hamiltonian}.
In our carbon demonstration below, we retain only the latter term to ensure that BSR and RMI use the same SO operator.

The interaction with the laser field, in the length form, is given by
\begin{equation}
 H_{\mathrm{int}}(t)=\mathbf{E}(t)\cdot\sum_{i=1}^{N+1}\mathbf{r}_i,
 \label{eq:laser}
\end{equation}
where the comparisons below use linear polarization along $z$, $\mathbf{E}(t)=E(t)\hat{\mathbf{z}}$.

\mbox{BSR--RMT} is not ultimately limited to the one-body SO term; additional Breit--Pauli (BP) contributions included in the BSR atomic data~\cite{ZatsarinnyFischer2000,Zatsarinny2006} can be transferred without further adjustment to the utilities presented here.
The comparisons in the present work include only the one-body SO term in order to conduct a benchmark comparison with the established \mbox{RMI--RMT} treatment~\cite{Wragg2020}.
SO effects result in energy and wave-function shifts in the target and neutral states, leading to adjustments to the structure quantities, along with opening additional pathways via mixing.
While the outer-region channels maintain the SO-split target thresholds in RMT, the propagation of the outer electron is nonrelativistic.
Thus, extending the calculation to other BP terms requires that those terms be handled consistently with the outer-region treatment.

\section{Interface Utilities and Usage}\label{sec:interface}
After preparing the target states, along with their necessary phase convention, as described in Section~\ref{sec:target-states}, BSR constructs the inner-region $(N+1)$-electron states using the appropriate Hamiltonian.
Following BSR, the interface utilities listed in Table~\ref{tab:utilities} prepare the Hamiltonian, dipole, and B-spline input data required by RMT.
The Hamiltonian utility needs to be run before the dipole utility, since the latter uses long-range coefficients from the \code{H} output file of the former.
Table~\ref{tab:utilities} builds on the descriptions of the $LS$ utilities in Ref.~\cite{DelValle2025}.

\begin{table}[!htb]
\caption{Interface utilities and their output files.
The two spline utilities are common to the $LS$ and $jK$ preparations.}
\label{tab:utilities}
\begingroup
\renewcommand{\arraystretch}{1.15}
\renewcommand{\tabularxcolumn}[1]{p{#1}}
\begin{tabularx}{\linewidth}{@{}>{\raggedright\arraybackslash}p{0.31\linewidth}>{\raggedright\arraybackslash}X@{}}
\toprule
Utility & Output and purpose \\
\midrule
\code{bsr_Splinedata}
& Generates \code{Splinedata} from \code{knot.dat}, containing information about the B-splines such as knot sequence, spline order, number of intervals. \\
\addlinespace[0.6em]
\code{bsr_Splinewaves}
& Creates \code{Splinewaves}, which contains the channel components of the inner-region eigenstates generated by \code{bsr_hd3}, in the B-spline basis. \\
\addlinespace[0.6em]
\code{bsr_H_jk}
& Assembles \code{H} from the \texttt{h.nnn} files. This file passes energies, channel information, long-range coefficients, and boundary amplitudes to RMT in the $jK$ coupling scheme. \\
\addlinespace[0.6em]
\code{bsr_d_jk}
& Creates \code{D00} and \texttt{D\#\#\#}, which contain reduced dipole matrix elements and the associated angular information for dipole transitions between pairs of $N$-electron target states and between pairs of $(N+1)$-electron R-matrix eigenstates. \\
\bottomrule
\end{tabularx}
\endgroup
\end{table}

For completeness, further modifications to BSR were made solely to improve numerical accuracy in the comparisons.
The radial integration routine was modified for certain \hbox{B-spline} grids already supported by BSR, and the precision used to save the target-mixing coefficients from MCHF was increased.
The integration change, along with other similar modifications needed in the helium comparisons in Section~\ref{sec:helium}, is described in Appendix~\ref{app:helium-numerics}.

\section{Results and Validation}\label{sec:results}
\subsection{Helium: structure and high-order harmonic generation}\label{sec:helium}
Two-electron helium provides a simple test of the $jK$ interface in a light system where we expect it to approach the nonrelativistic limit.
We consider examples with one and three nonrelativistic He$^+$ targets: $1\mathrm{s}\,{}^2\mathrm{S}$ alone, or $1\mathrm{s}\,{}^2\mathrm{S}$, $2\mathrm{s}\,{}^2\mathrm{S}$, and $2\mathrm{p}\,{}^2\mathrm{P}^{\mathrm{o}}$.
In $jK$ coupling, the ``three-target model'' is actually four targets: $1\mathrm{s}\,{}^2\mathrm{S}_{1/2}$, $2\mathrm{s}\,{}^2\mathrm{S}_{1/2}$, $2\mathrm{p}\,{}^2\mathrm{P}^{\mathrm{o}}_{1/2}$, and $2\mathrm{p}\,{}^2\mathrm{P}^{\mathrm{o}}_{3/2}$.
In the one-target ground-state expansion, the He ground state contains only the $1\mathrm{s}$ orbital. 
Therefore, the one-body SO interaction vanishes, though SO mixes the excited $1\mathrm{s}n\mathrm{p}\,{}^1\mathrm{P}^{\mathrm{o}}_1$ and $1\mathrm{s}n\mathrm{p}\,{}^3\mathrm{P}^{\mathrm{o}}_1$ states, causing changes in the dynamics.
In the three-target model, on the other hand, SO mixing between the $^1\mathrm{S}_0$ state and even-parity $^3\mathrm{P}_0$ states produces a small second-order shift in the $J^\pi=0^{\mathrm{e}}$ ground-state energy.

\Needspace{4\baselineskip}
Table~\ref{tab:helium-structure} compares the SO-induced change in energy levels as well as in both the $LS$-allowed and intercombination SO-enabled oscillator strengths.
Here, we define \hbox{$\Delta X=X_{\SO}-X_{\NR}$}, using the same $jK$ model (either BSR or RMI) with one-body SO effects either switched on or off.
In Table~\ref{tab:helium-structure}, for both BSR and RMI we use 71 order-eight B-splines and an R-matrix radius of $30$~a.u.
Since the singlet-triplet oscillator strength vanishes without SO, $\Delta f=f_{\SO}$ in this case.

In the three-target state comparison, we increased the number of significant figures retained in the MCHF energies and adjusted the radial integration grid to match between BSR and RMI. 
We further improved the precision of the energies and eigenvectors by using higher-precision arithmetic in the diagonalization, as described in Appendix~\ref{app:helium-numerics}.
With these minimal changes, the SO-induced changes to the energies and oscillator strengths are in excellent agreement.

\begin{table}[!htb]
\caption{SO-induced changes $\Delta X=X_{\SO}-X_{\NR}$ to the energies $E$ (a.u.) and dimensionless length-form oscillator strengths $f$ in helium.
For both models, SO and NR values are obtained using $jK$ coupling with the SO interaction switched on and off, respectively.
Oscillator strengths are from the $1\mathrm{s}^2\,{}^1\mathrm{S}_0$ ground state to the indicated $J^\pi=1^{\mathrm{o}}$ levels. }
\label{tab:helium-structure}
\begin{adjustwidth}{-\extralength}{0cm}
\centering
\begingroup
\setlength{\tabcolsep}{2pt}
\renewcommand{\arraystretch}{1.12}
\sisetup{table-format=+1.3e-2,retain-explicit-plus=true,table-number-alignment=center}
\begin{tabular*}{\linewidth}{@{\extracolsep{\fill}}lSSSS@{}}
\toprule
 & \multicolumn{2}{c}{One target} & \multicolumn{2}{c}{Three targets}\\
\cmidrule(lr){2-3}\cmidrule(l){4-5}
Quantity & {BSR} & {RMI} & {BSR} & {RMI}\\
\midrule
\(\Delta E(1s^2\,{}^1S_0)\) & 0 & 0 & -4.789e-11 & -4.789e-11\\
\(\Delta E(1s2p\,{}^3P_1^o)\) & -1.722e-06 & -1.722e-06 & -1.804e-06 & -1.804e-06\\
\(\Delta E(1s2p\,{}^1P_1^o)\) & +2.768e-10 & +2.761e-10 & +2.778e-10 & +2.778e-10\\
\(\Delta f({}^1P_1^o)\) & -1.403e-08 & -1.403e-08 & -1.330e-08 & -1.330e-08\\
\(\Delta f({}^3P_1^o)\) & +1.511e-08 & +1.511e-08 & +1.429e-08 & +1.429e-08\\
\bottomrule
\end{tabular*}
\endgroup

\end{adjustwidth}
\end{table}

For a dynamic comparison, we use the three-target structure expansion described above and the pulse details (15 cycles of linearly polarized light with a $\sin^2$ field envelope, a central wavelength of $248.6\,\mathrm{nm}$, and a peak intensity of $10^{14}\,\mathrm{W\,cm^{-2}}$) considered in our earlier study comparing HHG calculations in helium using several codes~\cite{Bondy2024}.
Here, we use $L_{\max}=9$ and $J_{\max}=9$ for NR and SO-on calculations, respectively.
A common structure, including the same angular cutoff, facilitates a controlled comparison of the BSR and Belfast R-matrix input data.
To test $jK$ dynamics, BSR and RMI use independently selected inner-region radial grids with 71 and 80 B-splines, respectively, with a common inner-region radius of $30$~a.u.
The three-target structure comparison in Table~\ref{tab:helium-structure} employs grids that were matched exactly to remove tiny differences that could emerge due to the radial basis in order to compare very small SO-induced shifts.
A time step of $0.001$~a.u. was used for propagation in RMT, along with an outer-region grid spacing of $0.08$~a.u., for a duration of $T=514.3$~a.u.
At the chosen angular cutoffs, the dipole responses and spectra are converged with respect to the numerical B-spline parameters along with time step and grid size.

\Needspace{4\baselineskip}
For both atoms we consider, the dipole response along the polarization axis is
\begin{equation}
 d(t)=-\left\langle\Psi(t)\left|\sum_{i=1}^{N+1}z_i\right|\Psi(t)\right\rangle.
 \label{eq:dipole}
\end{equation}
The helium HHG spectrum is obtained from the dipole acceleration,
\begin{equation}
 S_{\mathrm{HHG}}(\omega)=\left|\int_0^T \ddot d(t)e^{-i\omega t}\,\mathrm{d}t\right|^2,
 \label{eq:hhg-spectrum}
\end{equation}
where $\ddot d(t)$ is obtained by numerically differentiating the dipole response.

\begin{figure}[!htb]
\centering
\includegraphics[width=\linewidth]{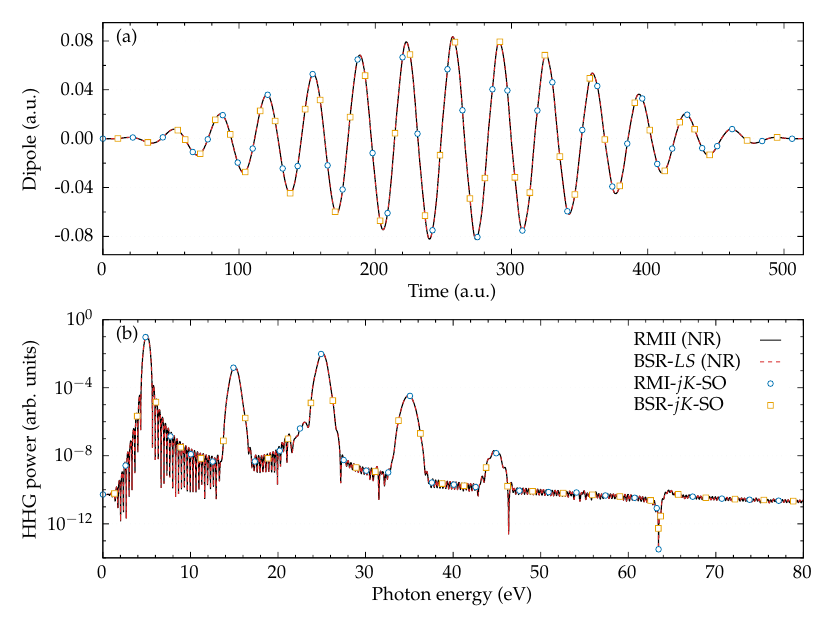}
\caption{Helium dipole response (a) and HHG spectrum (b), per Eq.~\eqref{eq:hhg-spectrum}, are shown for three-target BSR and Belfast R-matrix input into RMT.
The NR curves use $LS$ coupling, while the SO curves employ $jK$ coupling.
The pulse parameters are from Ref.~\cite{Bondy2024} and are given in the text.
\label{fig:helium-overlays}}
\end{figure}

Figure~\ref{fig:helium-overlays} compares \mbox{BSR-$LS$} and \mbox{RMII-$LS$}, with no SO-effect, alongside \mbox{BSR-$jK$} and \mbox{RMI-$jK$}, where SO is included.
The dipole responses and HHG spectra are in close agreement between BSR and the Belfast codes in both cases.
While these helium calculations compare the dynamical response in a limiting case where we expect close agreement with the NR case, the carbon example in the ensuing Section~\ref{sec:carbon} more stringently tests much larger changes induced by SO.

\subsection{Carbon: multiconfiguration response with and without SO}\label{sec:carbon}
We use carbon to perform a multiconfigurational comparison of \mbox{BSR--RMT} with (i)~\mbox{RMII--RMT} (in $LS$ coupling without SO), and (ii)~\mbox{RMI--RMT} (in $jK$ coupling with SO included).

\subsubsection{Atomic model and initial state}
The C$^+$ ground $^2\mathrm{P}^{\mathrm{o}}$ term contains the $1\mathrm{s}^22\mathrm{s}^22\mathrm{p}$ and $1\mathrm{s}^22\mathrm{p}^3$ configurations, which share a common set of target orbitals.
In this NR case, the weight of the $1\mathrm{s}^22\mathrm{p}^3$ configuration is $c^2=4.8\%$.
The $^2\mathrm{P}^{\mathrm{o}}$ term is the only term retained in $LS$-coupling, whereas the $jK$ simulations include its $J_t=1/2$ and $3/2$ levels.

As mentioned above, the target states and the orbitals from which they are constituted are generated using the MCHF program.
In order to compare the codes as closely as possible, we attempt to use the same orbitals. 
In the $LS$ comparison, the MCHF orbitals are used by BSR directly, while Slater-function fits that closely reproduce them (relative root-mean-square differences below $10^{-6}$) are employed in RMII.
For the $jK$ comparison, the MCHF orbitals are again supplied to RMI, where they are represented in a B-spline basis and orthogonalized.
We then supply these final orbitals to BSR.
BSR takes the target energies and mixing coefficients generated in MCHF and changes the latter to FR convention according to Eq.~\eqref{eq:coefficient-phase}.
RMII and RMI take the aforementioned orbitals and proceed with their own target state structure routines.

In this weak-field response comparison, the $(N+1)$-electron $LS$ expansion of neutral carbon includes the $^3\mathrm{P}^{\mathrm{e}}$, $^3\mathrm{P}^{\mathrm{o}}$, $^3\mathrm{D}^{\mathrm{e}}$, and $^3\mathrm{D}^{\mathrm{o}}$ symmetries, with continuum partial waves $\mathrm{p}$; $\mathrm{s},\mathrm{d}$; $\mathrm{p},\mathrm{f}$; and $\mathrm{d}$, respectively.
In $jK$ coupling, we include the two $\mathrm{p}$-wave channels in the $J^\pi=0^+$ symmetry and five channels (two $\mathrm{s}$-waves and three $\mathrm{d}$-waves) in the $1^-$ sector.

The initial-state expansion also includes the $1\mathrm{s}^22\mathrm{s}^22\mathrm{p}^2$ and $1\mathrm{s}^22\mathrm{p}^4$ bound configurations in the $^3\mathrm{P}^{\mathrm{e}}$ symmetry in $LS$ coupling, while in $jK$ coupling we include the $^3\mathrm{P}_0$ components of these configurations.
These are included as perturbers in BSR's \code{target} input files and as additional square-integrable configurations in RMII and RMI, specified in \code{C.inp} and \code{dstg2}, respectively.
Their mixing is determined in each code separately by diagonalizing the carbon Hamiltonian.
No additional bound configurations are included in the odd-parity sector.

The SO-on calculations begin in the lowest $J=0$, $M_J=0$ level, identified as the $^3\mathrm{P}_0$ ground state of neutral carbon.
In $LS$ coupling, the initial state is the $^3\mathrm{P}^{\mathrm{e}}$ ground term, and for comparison with the SO-on calculation, the $LS$ dipole response is calculated according to
\begin{equation}
 d_{\NR}(t)=\frac{d_{M_L=0}(t)+2\,d_{M_L=1}(t)}{3},
 \label{eq:average}
\end{equation}
which uses the fact that the $M_L=+1$ and $-1$ responses are equal for linear polarization along $z$.
The averaged dipole in Eq.~\eqref{eq:average} is then Fourier transformed to obtain the spectra.

The four calculations all use order-eight B-splines with an R-matrix radius of $30$~a.u. and an outer-region grid spacing of $0.08$~a.u.
The $LS$ calculations, BSR and RMII, employ the same grid with 56 splines; the two $jK$ calculations, BSR and RMI, use an identical grid with 71 splines generated on the exponential-knot sequence from RMI.
For each pair, the knot sequences are matched within numerical precision. 

\subsubsection{Pulse and dipole comparison}
The linearly polarized, four-cycle pulse has the form
\begin{equation}
 E(t)=E_0\sin^2\!\left(\frac{\pi t}{T_p}\right)
 \cos\!\left[\omega_0\left(t-\frac{T_p}{2}\right)\right],
 \qquad 0\leq t\leq T_p,
 \label{eq:pulse}
\end{equation}
and is otherwise zero.
The central frequency, $\omega_0=0.30$ a.u., was purposely chosen to be close to dipole-allowed absorption near $0.31$ a.u., to ensure a strong dipole response at relatively weak intensity.
The pulse duration is $T_p=8\pi/\omega_0 \simeq 83.776$ a.u., and the electric field amplitude is $E_0=0.0005336$ a.u., corresponding to a peak intensity of $10^{10}\,\mathrm{W\,cm^{-2}}$.
Following the pulse, field-free propagation continues to $T=500$ a.u.~to further test dynamical agreement.

The simulations in all cases were conducted with the unmodified RMT.
The $LS$ simulations used an order-20 propagation with $\Delta t=0.00125$ a.u., while in $jK$ calculations, order-24 propagation and $\Delta t=0.0003125$ a.u.~were chosen.
Convergence has been obtained with respect to the aforementioned time steps; in particular, in the SO-on case, the further change due to smaller time steps is much less than the SO-induced effect discussed below.

\begin{figure}[!b]
\centering
\includegraphics[width=\linewidth]{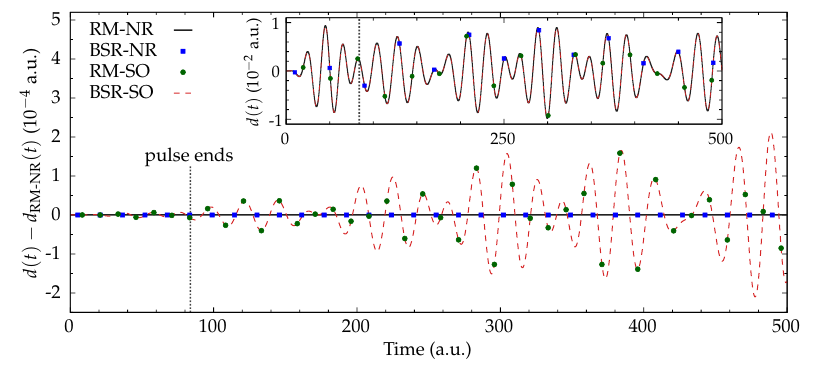}
\caption{Dipole response (including post-pulse propagation) in multiconfigurational carbon, starting in its ground state, both with and without SO effects. 
In $LS$-coupling, averaging is performed according to Eq.~\eqref{eq:average}.
The main panel compares differences in the responses compared with \mbox{RM-NR}, $d(t)-d_{\mathrm{RM\text{-}NR}}(t)$. 
The inset displays the full responses.
The grey vertical line indicates the end of the pulse at $T_p\simeq83.78$ a.u., which is defined in Eq.~\eqref{eq:pulse}.}
\label{fig:carbon-dipole}
\end{figure}

Figure~\ref{fig:carbon-dipole} displays the dipole response from Equation~\eqref{eq:dipole}.
In the legend, \mbox{RM-NR} is \mbox{RMII-$LS$}, \mbox{BSR-NR} is \mbox{BSR-$LS$}, \mbox{RM-SO} is \mbox{RMI-$jK$}, and \mbox{BSR-SO} is \mbox{BSR-$jK$} with SO included.
The inset shows the full dipole response, while the main panel exhibits small differences by comparing all responses to \mbox{RM-NR}.
The legend describes both the main panel and the inset. The RM-NR curve in the main panel is identically zero by construction and is nevertheless plotted as a visual reference.
The BSR-NR trace is indistinguishable from the zero line at this scale, demonstrating excellent agreement between the NR versions of the two codes.

The maximum absolute difference in the BSR and RM dipoles, divided by the maximum absolute RM response, is $3.8\times10^{-6}$ in the NR $LS$ case and $2.4\times10^{-6}$ with SO included.
These differences are several orders of magnitude smaller than the SO-induced change and facilitate the intended comparison.

While the coupling scheme should not affect the NR results, our $LS$ and $jK$ calculations used slightly different B-spline expansions and an angular description that is not identical.
To test the consequence of this, an SO-off, NR RMI calculation was performed in the $jK$ basis.
Comparing with the $LS$ result, we found that this difference is also much smaller than the SO-induced effect.

In particular, since we are near resonances, the dipole continues to oscillate strongly after the pulse, making the SO-induced difference even more visible.
The maximum NR--SO difference becomes 2.2\% of the peak dipole in the post-pulse period, compared with only 0.08\% during the pulse itself.
Separate static SO-included structure comparisons between the $^3\mathrm{P}_J$ levels (with $J=0,1,2$) show excellent agreement ($6\times10^{-10}$ a.u.\ or better) between BSR and RMI in the SO-induced energy shifts.

\subsubsection{Dipole power spectrum}
\begin{figure}[!b]
\centering
\includegraphics[width=\linewidth]{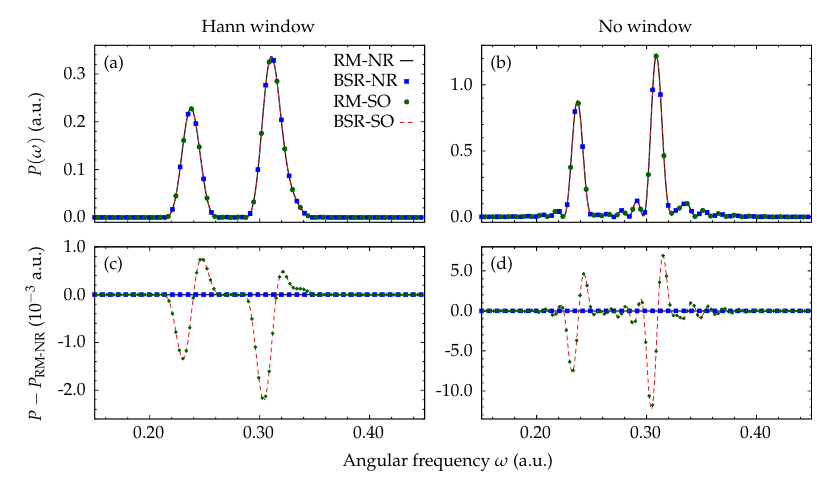}
\caption{Dipole power spectra in the indicated carbon models using the Hann window (a,c) and without windowing (b,d), corresponding to the full dipole responses in Figure~\ref{fig:carbon-dipole}.
Panels (a,b) display the full spectra while panels (c,d) contain the difference from \mbox{RM-NR}.}
\label{fig:carbon-spectrum}
\end{figure}

Figure~\ref{fig:carbon-spectrum} displays the corresponding dipole spectra
\begin{align}
 D_w(\omega)&=\int_0^T w(t)d(t)e^{-i\omega t}\,\mathrm{d}t,
 & P(\omega)&=|D_w(\omega)|^2.\label{eq:spectrum}
\end{align}

\noindent To illustrate the effect of windowing, and that agreement persists both with and without windowing, we compare the unwindowed result, $w(t)=1$, with the Hann window
\begin{equation}
 w(t)=\sin^2\!\left(\frac{\pi t}{T}\right),
 \qquad 0\leq t\leq T,
 \label{eq:window}
\end{equation}
taking the entire $T=500$ a.u.~as the time interval.
The Hann window reduces spectral leakage that results from the signal truncation, though it leads to broadening the spectral features~\cite{Harris1978}.

The longer observation window enables dipole-allowed excited states near $0.24$ and $0.31$ a.u.\ to be distinguished in the spectra, via post-pulse oscillations.
In our carbon model, the first corresponds to the lowest $^3\mathrm{P}^{\mathrm{o}}$ state, while the higher energy includes the lowest $^3\mathrm{D}^{\mathrm{o}}$ state and higher $^3\mathrm{P}^{\mathrm{o}}$ excitations (these contribute to the corresponding $J^\pi=1^{\mathrm{o}}$ states).
Over an observation window lasting only the pulse duration itself, the lower-energy feature is less clearly resolved; however, the extended propagation makes it easier to distinguish.
Note that these features are present in both the NR and SO calculations and that they do not represent the splitting of an SO doublet.
Though it would be possible to observe SO-induced energy shifts of dipole-allowed states in this fashion, a much longer propagation, of order $T\sim10^5$ a.u., would be needed.

\Needspace{14\baselineskip}
\section{Conclusions and Outlook}\label{sec:conclusions}
The present work extends the \mbox{BSR--RMT} interface to include multiconfiguration target states and $jK$ coupling, facilitating the inclusion of the one-body SO interaction.
These extensions allow the \mbox{BSR--RMT} interface to treat a broader range of systems already accessible through the Belfast R-matrix to RMT pipeline.
As before, the interface supplies Hamiltonian, dipole, and spline data directly for use in an unmodified version of RMT.
We performed structure tests in helium showing outstanding agreement between BSR and RMI for the very small SO-induced changes to both energy and oscillator strengths.
Close dynamical agreement between the predictions from the two codes, both with and without SO effects, was also demonstrated in helium by comparing the dipole responses and HHG spectra.
In carbon, matched multiconfigurational target descriptions likewise yield very similar dipole responses and corresponding spectra in the two codes.
In particular, the SO-induced change in these curves is much larger than in helium and is closely reproduced by both BSR and RMI.

The core vision of this \mbox{BSR--RMT} program~\cite{Schneider2022,DelValle2025,BondyBartschat2026} was to combine the flexibility of the target description in BSR with the highly successful RMT code for simulations in multielectron atoms.
The independently optimized target states in BSR use nonorthogonal orbital sets that in stationary collision, electron impact, and photoionization calculations have enabled accurate target descriptions and compact configuration expansions.
In heavier, more complex and possibly open-shell atoms, in particular, this flexibility is expected to provide an advantage over conventional methods.
While the present results compared computationally tractable test cases to benchmark the \mbox{BSR--RMT} interface, it is now ready and will be used to tackle the aforementioned more challenging systems.

\authorcontributions{
A.T.B. and J.C.d.V. performed the calculations and analyzed and interpreted the results.
A.T.B. and J.C.d.V. contributed to the development of the software used in this project.
K.B. and K.R.H. envisioned the project and provided guidance with the interpretation of results.
A.T.B. prepared the initial manuscript.
All authors contributed to reviewing and editing the manuscript.}

\funding{This work was supported by the National Science Foundation under grant Nos. \hbox{PHY-2110023} and PHY-2408484. Computational resources were provided through ACCESS project MCA08X034 on Stampede3 at the Texas Advanced Computing Center and Expanse at the San Diego Supercomputer Center, and through Frontera allocation PHY20028 at the Texas Advanced Computing Center.}

\dataavailability{
The interface codes, described in Section~\ref{sec:interface}, along with example inputs and run scripts, will be made available at the time of publication.
The following are available from the corresponding author upon request: (i) the BSR code (with FR convention), (ii) modified MCHF routines to facilitate SO-only target states needed for the present comparison, (iii) the modified routines used to improve precision for the structure comparison in Section~\ref{sec:helium} (these are described in more detail in Appendix~\ref{app:helium-numerics}), and (iv) all datasets included in the paper.}

\acknowledgments{We thank Hugo van der Hart and Andrew Brown for helpful exchanges concerning the use of the Belfast R-matrix codes and RMT.}

\conflictsofinterest{The authors declare no conflicts of interest.}

\appendixtitles{yes}
\appendixstart
\appendix
\Needspace{7\baselineskip}
\section{Numerical accuracy in the helium comparisons}\label{app:helium-numerics}
Comparing the extremely small SO-induced changes in helium requires higher numerical accuracy than is needed to establish agreement for either the energies themselves or the overall dynamical responses/spectra.
Here we describe how we improved the numerical precision to facilitate these comparisons, without changing any operators in BSR or the Belfast R-matrix codes, and without changing anything in RMT propagation.

In the three-target structure comparison, the double-precision elements in the Hamiltonian matrix were kept, as was the overlap matrix in BSR.
Following standard diagonalization, the residual error in the matrix eigenvalue equations was evaluated in higher-precision (60-digit) arithmetic, allowing these equations to be solved more accurately for the energies and eigenvectors.
The dipoles and oscillator strengths were then calculated using these more accurate eigenvalues and eigenvectors, carrying the higher-precision arithmetic through intermediate calculations.

We further reduced the MCHF rounding error of the target energies by retaining seventeen decimal places rather than the usual eight and ensured that these digits were kept throughout target preparation.
As part of its numerical integration scheme, the original grid in BSR allows some Coulomb integrals to be calculated not directly, but by scaling the integrals obtained on other intervals.
To facilitate a 1-to-1 comparison with RMI, we calculate the integrals on each successive knot interval.
The matrix elements on the original BSR grid remain unchanged.
With these minor, precision-only modifications to the code, all five (very small) quantities in the three-target helium model shown in Table~\ref{tab:helium-structure} agree to four significant figures.
In one-target helium, these structure quantities agree closely without the additional precision.

Despite the good agreement for the full time-dependent dipoles/spectra in three-target helium shown in Fig.~\ref{fig:helium-overlays}, we performed one more test, checking the sensitivity of the SO-induced dipole change to the increased numerical precision described above.
The previously described structure comparison used the higher-precision diagonalization in both codes, but for this dynamical test the changes were only made in RMI.
Within RMI, we reorthogonalized the eigenvectors and then rediagonalized the Hamiltonian with higher-precision (in this case 31-digit) arithmetic, and used the resulting more precise atomic structure as input for a truncated propagation up to 300 a.u.
The total time-dependent RMI dipole changed at most by $2\times10^{-9}$~a.u., while the peak dipole response is $0.08$~a.u.
For both BSR and RMI, we then isolated the SO-induced change by taking the difference between the SO-on and SO-off dipoles.
In this interval up to 300 a.u., improving RMI by the aforementioned changes to the diagonalization routine reduced the maximum difference between the SO-induced dipole changes obtained with \mbox{BSR--RMT} and \mbox{RMI--RMT} from $2.05\times10^{-9}$ to $4.8\times10^{-12}$~a.u.
The BSR propagation was unchanged and did not include the precision adjustments described above for the structure comparison in Table~\ref{tab:helium-structure}.

In carbon, on the other hand, the SO-induced dipole change is 2.17\% of the peak, whereas the BSR--RM dipole differences are at the ppm level.
Increased precision is not a general requirement for using the interface and was unnecessary to resolve the SO effect in the present carbon example.

\begin{adjustwidth}{-\extralength}{0cm}
\reftitle{References}
\bibliography{references}
\end{adjustwidth}
\end{document}